\documentclass[useAMS,usenatbib]{mn2e}
\usepackage{rotating}
\usepackage{lscape}
\title[Chemical composition of evolved stars in NGC\,2506]
      {Chemical composition of evolved stars in the open cluster NGC\,2506\thanks{Based on observations collected at ESO telescopes under 
programmes 65.N-0286 and in part 169.D-0473}}

\author[\v{S}ar\={u}nas Mikolaitis et al.]
       {\v{S}ar\={u}nas Mikolaitis,$^{1}$\thanks{E-mail: sarunas.mikolaitis@tfai.vu.lt}
       Gra\v{z}ina Tautvai\v sien\. e,$^{1}$ 
       Raffaele Gratton,$^{2}$
       Angela Bragaglia$^{3}$ 
       \newauthor and Eugenio Carretta$^{3}$\\ 
$^{1}$Institute of Theoretical Physics and Astronomy, Vilnius University, Go\v{s}tauto 
12, Vilnius 01108, Lithuania\\
       $^{2}$INAF - Osservatorio Astronomico di Padova, Vicolo dell'Osservatorio 5, I-35122 Padova, Italy\\
       $^{3}$INAF - Osservatorio Astronomico di Bologna, Via Ranzani 1, I-40127 Bologna, Italy}
       
\begin{document}

\date{Accepted to MNRAS May 19 2011}

\pagerange{\pageref{firstpage}--\pageref{lastpage}} \pubyear{2011}

\maketitle

\label{firstpage}

\begin{abstract}
In this study we present abundances of  $^{12}$C, $^{13}$C, N, O and up to 26 other chemical elements 
in two first ascent giants and two core-helium-burning `clump' stars of the open cluster NGC\,2506. 
Abundances of carbon were derived using the ${\rm C}_2$ Swan (0,1) band head at
5635.5~{\AA}. The wavelength interval 7940--8130~{\AA}, with strong CN features, was analysed 
in order to determine nitrogen abundances and carbon isotope ratios.  
The oxygen abundances were determined from the [O\,{\sc i}] line at 6300~{\AA}.  
NGC\,2506 was found to have a mean ${\rm [Fe/H]}=-0.24\pm 0.05$ (standard deviation). 
Compared with the Sun and
other dwarf stars of the Galactic disc, mean abundances in the investigated
clump stars  suggest that carbon is depleted by about 0.2~dex, nitrogen is
overabundant by about  0.3~dex and other chemical elements have abundance ratios close to solar. 
The C/N and $^{12}{\rm C}/^{13}{\rm C}$ ratios are lowered to $1.25\pm 0.27$ and $11\pm 3$, 
respectively. 
\end{abstract}


\section{Introduction}

This work is continuing our efforts in studying the detailed chemical composition of 
stars in open clusters (Tautvai\v{s}ien\.{e} et al.\ 2000, 2005; Mikolaitis et al.\ 2010, 2011).
Open clusters have several very valuable features: they ensemble stars with a common age, initial 
chemical composition, and distance; they are distributed across the entire Galactic disc and span 
a wide range of ages; stars inside a cluster are in different evolutionary stages. This makes 
open clusters valuable sources of information in many areas of astrophysical investigations. 
 
In this work, our target of investigation is the open cluster NGC\,2506 ($\alpha_{2000}=08^{h}0.02^{m}, 
\delta_{2000}=-10^{\circ}46.2^{\prime}; l = 230.564^{\circ}, b = +09.935^{\circ}$). 
The Galactic orbit of this cluster was determined by Carraro \& Chiosi (1994). It was found that the orbit of 
NGC\,2506 has a small eccentricity ($e=0.03$) and epicyclical amplitude ($\Delta R=0.84$~kpc), suggesting 
that it has not moved far away from the site of formation. The orbit remains confined at radial 
distances between 10.7 and 11.6~kpc, and along the $z$-direction does not extend beyond 0.6~kpc. 
     
Since the first studies by van der Bergh \& Sher (1960) and Purgathofer (1964), NGC\,2506 has quite many references in 
the literature. Relative proper motions for 724 stars in the region of NGC\,2506 have been determined and 
probabilities of membership derived by Chiu \& van Altena (1981). According to the recent studies, 
NGC\,2506 is a mildly elongated cluster containing about 1090 stars (Chen, Chen \& Shu 2004), its generally accepted 
age is $t=1.7$~Gyr (Marconi et al.\ 1997), the turn-off mass $M=1.69~M_{\odot}$ (Carretta et al.\ 2004, hereafter C04), 
its Galactocentric and Heliocentric distances are $R_{gc}=10.38$~kpc and $d=3.26$~kpc respectively (Bragaglia \& Tosi 2006).

A high resolution spectroscopic observation of NGC\,2506 has been obtained already 30 years ago 
by Geisler (1984). For the star 22012,  a surprisingly low metallicity ([Fe/H]\footnote{In this paper we use the 
customary spectroscopic notation
[X/Y]$\equiv \log_{10}(N_{\rm X}/N_{\rm Y})_{\rm star} -
\log_{10}(N_{\rm X}/N_{\rm Y})_\odot$}$=-0.67$) was determined 
and [Mg/Fe] and [Si/Fe] were found to be rather large (0.8~dex).  
Metallicity determinations using Washington system photometry were also presented for this cluster in the 
same paper, with value of $-0.93$~dex using the $M-T_1$ colour index and of $-0.51$~dex using $C-M$.        
Later photometric observations of this cluster indicated quite a low metallicity as well. A value of
${\rm [Fe/H]}=-0.55$ relative to 
the Hyades was determined from $UBV$ and DDO photoelectric photometry and $B$ and $V$ photographic photometry by 
McClure, Twarog \& Forrester (1981). Using the same observational data and a new calibration, Piatti, 
Claria \& Abadi (1995) have determined ${\rm [Fe/H]}=-0.48$. From Washington photoelectric photometry  
${\rm [Fe/H]}=-0.58$ was determined by Geisler, Claria \& Minniti (1992).        
However, a higher metal abundance of $-0.20$~dex was obtained from high resolution spectroscopy by C04, 
as averaged from two stars. 
 
In this work, we analysed four stars in NGC\,2506 observed by C04. Using a carefully selected line-list, 
we redetermined the main atmospheric parameters, derived abundances of $^{12}{\rm C}$, $^{13}{\rm C}$ and 
nitrogen, which will contribute to the expansion of a database on abundances of mixing-sensitive 
chemical elements in open clusters. The degree of mixing in stars largely depends on their masses and 
metallicities. The available number of open clusters investigated so far is far from sufficient in order to 
make reliable models of stellar evolution (Charbonnel \& Lagarde 2010, Mikolaitis et al.\ 2011, and 
references therein).       

The data on abundances of heavier chemical 
elements in NGC\,2506 will be used to study time evolution of abundances in the Galactic disc 
within the Bologna Open Cluster Chemical Evolution (BOCCE) study 
(Bragaglia \& Tosi 2006; Carretta, Bragaglia \& Gratton 2007).

  
\input epsf
\begin{figure}
\epsfxsize=\hsize 
\epsfbox[-5 -10 600 550]{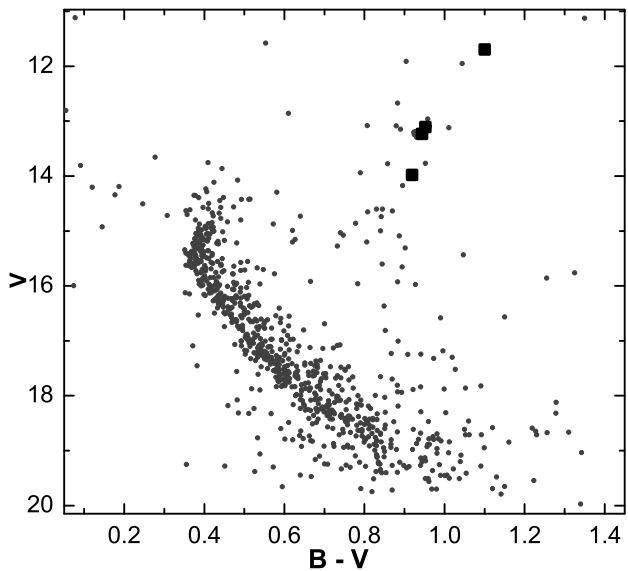} 
\caption{The colour-magnitude diagram of the open cluster
NGC\,2506. The stars investigated in this work are indicated by
the filled squares. The diagram is based on $UBVRI$ photometry
 by Marconi et al.\ (1997).} 
\label{fig1}
\end{figure}
 

\section{Observations and method of analysis}
The spectra of four stars (NGC\,2506 438, 443, 456 and 459) were obtained with 
the Fiber-fed Extended Range Optical 
Spectrograph (FEROS) mounted at the 1.5~m telescope in La Silla (Chile) by C04. 
Two stars (438 and 443) belong to the red clump, the star 456  
is a first-ascent giant, and the star 459 is an red giant branch (RGB)-tip giant (see Fig.~1 for the 
colour-magnitude diagram of NGC\,2506 with these stars indicated). 
Small portions of the spectra are presented in Fig.~2. The resolving power is 
$R=48\,000$ and the full wavelength range is $\lambda\lambda$ 3700--8600\,{\AA}. 
The spectrum of RGB-tip star 459 has the highest signal-to-noise ratio $S/N=110$, while 
S/N of the spectrum of star 456, which lies below the clump, is only 35.  
The log of observations and 
S/N for individual stars are presented in the paper by C04.

\input epsf
\begin{figure}
\epsfxsize=\hsize 
\epsfbox[20 20 300 220]{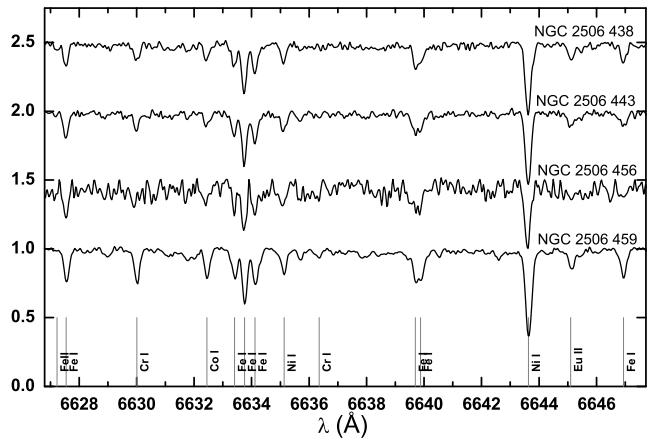} 
    \caption {Samples of stellar spectra of all programme stars in NGC\,2506. An offset 
 of 0.5 in relative flux is applied for clarity.}
    \label{Fig2}
  \end{figure}

Since the S/N ratios of stellar spectra observed in this cluster were much lower than in the other 
open clusters observed by C04 and investigated in our previous papers (Mikolaitis et al.\ 2010, 2011), 
we decided to try to redetermine the main atmospheric parameters of stars in NGC\,2506 using more 
strictly selected lines. 
   
Effective temperature, surface gravity and microturbulent velocity were derived using traditional 
spectroscopic criteria.
The preliminary values of effective temperatures were derived using $(B-V)_0$ colour indices and the  
temperature calibration by Alonso, Arribas \& Mart\,{i}nez-Roger (2001).
The interstellar reddening value $E_{B-V} = 0.07$, as averaged from four determinations (Marconi et al.\ (1997); 
Schlegel, Finkbeiner \& Davis 1998; Kim et al.\ 2001; and Dias et al.\ 2002), was taken into account.   
 
The spectroscopic effective temperatures were derived by minimizing a slope of the abundances obtained from neutral 
Fe\,{\sc i} lines with respect to the excitation potential. Differences between the photometric and spectroscopic 
effective temperatures did not exceed 100~K.
The gravities (log~$g$) were derived by forcing measured neutral 
and ionised iron lines to 
yield the same [Fe/H] value by adjusting the model gravity.  
The microturbulent velocities were determined by forcing Fe\,{\sc i} abundances to be independent of the equivalent 
widths of lines. 

The {\sc atlas} models with overshooting were used for an analysis of the spectra. 
Spectral lines were restricted to the spectral range 5500--8000\,{\AA} in order to minimize 
problems of line crowding and difficulties in the continuum tracing in the blue region. After the careful selection, 
the number of Fe\,{\sc i} lines initially analysed by C04 was reduced from 137--102 to 47--49 in the brighter 
stars and from 83 to 35 in the fainter star NGC\,2506 456. The number of Fe\,{\sc ii} lines was reduced from 13--10 
to 6--5, respectively. This allowed us to minimize the scatter and increase the accuracy of the abundance and stellar 
atmospheric parameter determinations. The determined atmospheric parameters and iron abundances for the observed stars 
in NGC\,2506 are presented in Table~1. 

   \begin{table*}
\caption[]{Adopted atmospheric parameters for observed stars in NGC~2506.
}
\label{Param}
      \[
         \begin{tabular}{rcccccccccc}
            \hline
            \noalign{\smallskip}
Star$^*$  & $V$ & $B-V$  &  $T_{\rm eff}$ & log~$g$ & $v_t$ & [Fe/H] &  $\sigma_{\rm Fe I}$ &$n_{\rm Fe I}$&$\sigma_{\rm Fe II}$&$n_{\rm Fe II}$\\
          & (mag) & (mag) &     (K)        &       & (km s$^{-1}$) &  \\
\hline
 438 &13.234  & 0.944 & 5050 & 2.64 & 1.5 & $-0.18$ & 0.08 & 49 & 0.07& 5\\
 443 &13.105  & 0.952 & 5050 & 2.60 & 1.6 & $-0.24$ & 0.07 & 47 & 0.06& 5\\
 456 &13.977  & 0.919 & 4960 & 2.96 & 1.8 & $-0.23$ & 0.06 & 45 & 0.05& 5\\
 459 &11.696  & 1.100 & 4660 & 2.16 & 1.7 & $-0.29$ & 0.06 & 49 & 0.05& 6\\
                \noalign{\smallskip}
            \hline
         \end{tabular}
      \]
$^*$ Star numbers, $V$ and $B-V$  from Marconi et al.\ (1997) 
   \end{table*}

In this work we used the same method of analysis as in Mikolaitis et al.\ (2010, 2011)
for the determination of abundances of other chemical elements. In the following, we give 
only some details about the spectral lines investigated. 

C abundances were derived from the ${\rm C}_2$ Swan 0 -- 1 band head at 5630.5~{\AA};
a fit of synthetic and observed spectra of this feature for the NGC\,2506\,459 spectrum is shown in Fig.~3.  
We derived oxygen abundances from synthesis of the forbidden [O\,{\sc i}] line at 6300.3~{\AA}. 
The $gf$ values for $^{58}{\rm Ni}$ and $^{60}{\rm Ni}$ isotopic line components, which blend the 
oxygen line, were taken from Johansson et al.\ (2003). In Fig.~4, we show an example of spectrum synthesis 
for the [O\,{\sc i}] line in NGC\,2506\,443. 
In spectra of the NGC\,2506 stars the [O\,{\sc i}] line was not contaminated 
by telluric lines.  

\input epsf
\begin{figure}
\epsfxsize=\hsize 
\epsfbox[-20 -20 620 580]{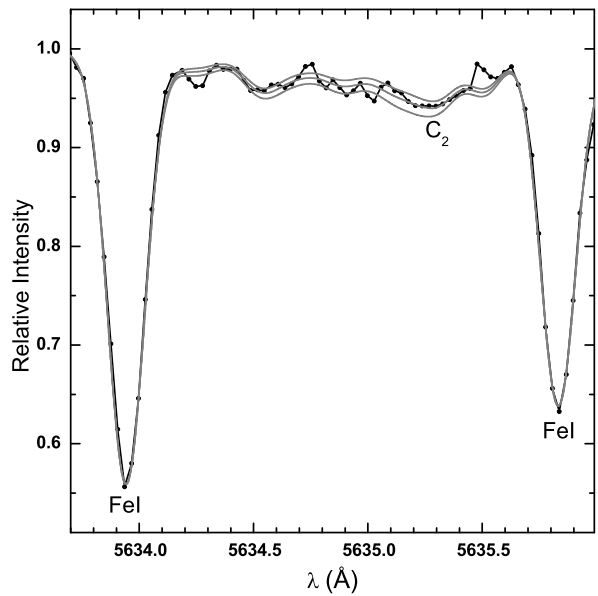} 
    \caption {A small region of NGC\,2506\,459 spectrum (solid black line with black dots) at the
${\rm C}_2$ Swan (0,1) band head 5635.5~{\AA}, plotted together with 
synthetic spectra (grey lines) with [C/Fe] values at $-0.19\pm 0.05$~dex. 
}
    \label{Fig3}
  \end{figure}

\input epsf
\begin{figure}
\epsfxsize=\hsize 
\epsfbox[-20 -20 620 470]{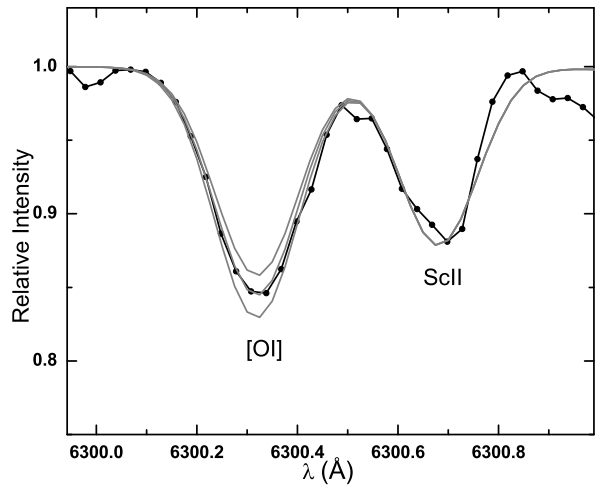} 
    \caption {A fit to the forbidden [O\,{\sc i}] line at 6300.1~{\AA} in 
NGC\,2506\,443. The observed spectrum is shown as a solid line with black dots. The synthetic
spectra with [O/Fe]$=0.02\pm 0.03$ are shown as solid grey lines.}
    \label{Fig4}
  \end{figure}

\input epsf
\begin{figure}
\epsfxsize=\hsize 
\epsfbox[-20 -20 620 470]{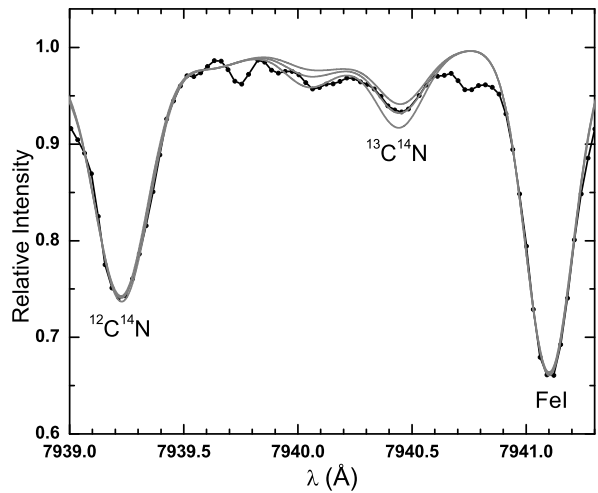} 
    \caption {
A small region of NGC\,2506\,459 spectrum (solid black line with black dots) 
    with $^{13}{\rm C}^{14}{\rm N}$ feature.
Grey lines show synthetic spectra with $^{12}{\rm C}$/$^{13}{\rm C}$ ratios equal to 10 (lowest line), 
14 (middle line) and 18 (upper line). 
}
    \label{Fig5}
  \end{figure}

The interval 7940 -- 8130~{\AA} containing strong $^{12}{\rm C}^{14}{\rm N}$ features was used for 
the nitrogen abundance analysis. 
Unfortunately, a $^{12}{\rm C}/^{13}{\rm C}$ ratio analysis using the $^{13}{\rm C}^{14}{\rm N}$ feature at 
8004.7~\AA\ was not possible because of blending by telluric lines. Thus, we selected 
another $^{13}{\rm C}^{14}{\rm N}$ feature at 7940.4~\AA (see Fig. 5 for the example). The molecular data for this CN band were 
provided by Bertrand Plez (University of Montpellier II). All $gf$ values were increased by $+0.021$~dex 
to fit the model spectrum to the solar atlas of Kurucz et al.\ (1984). 

The solar carbon and nitrogen abundances used in our work are log$A_{\rm C} = 8.52$ and 
log$A_{\rm N} = 7.92$ (Grevesse \& Sauval 2000).
  
\input epsf
\begin{figure}
\epsfxsize=\hsize 
\epsfbox[-20 -20 620 470]{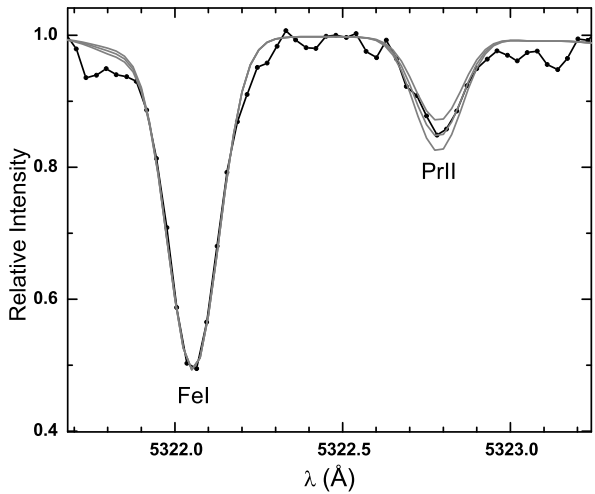} 
    \caption {A fit to the Pr\,{\sc ii} line at  5322.8~{\AA} in 
NGC\,2506\,438. The observed spectrum is shown as a solid line with black dots. The synthetic
spectra with [Pr/Fe]$=0.13\pm 0.10$ are shown as solid grey lines.}
    \label{Fig7}
  \end{figure}

\input epsf
\begin{figure}
\epsfxsize=\hsize 
\epsfbox[-20 -20 620 470]{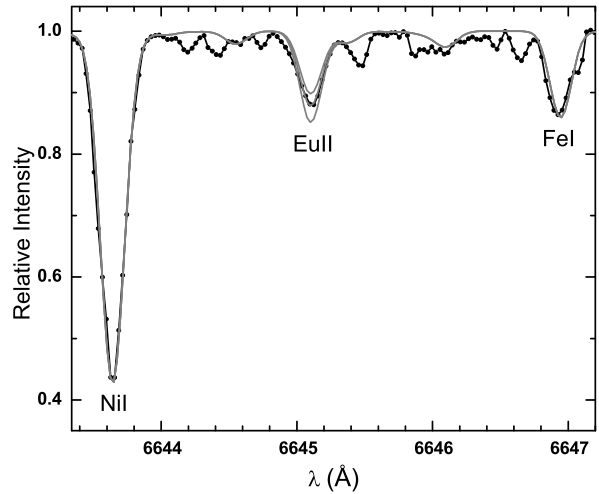} 
    \caption {A fit to the Eu\,{\sc ii} line at 6645.1~{\AA} in 
NGC\,2506\,438. The observed spectrum is shown as a solid line with black dots. The synthetic
spectra with [Eu/Fe]$=0.17\pm 0.10$ are shown as solid grey lines.}
    \label{Fig7}
  \end{figure}

Abundances of Na and Mg were determined with non local thermodynamic equilibrium (NLTE) taken into 
account as described by Gratton et al.\ (1999). The calculated corrections did not exceed 0.03~dex for Na\,{\sc i} 
and 0.09~dex for Mg\,{\sc i} lines.
Abundances of sodium were determined from equivalent widths of the Na\,{\sc i} lines 
at 6154.2 and 6160.8~{\AA}; magnesium from 
the Mg\,{\sc i} lines at 4730.0, 5711.1 and 6318.7~{\AA}; 
and that of aluminum from the Al\,{\sc i} lines at 7835.3 and 7836.1~{\AA}.

The determinations of copper, zirconium, yttrium, barium, lanthanum,
cerium, neodymium, praseodymium and europium abundances were performed by a spectral 
synthesis method. The copper abundances were derived using the Cu\,{\sc i} line at 5218.2~{\AA}, 
for which we adopted the hyperfine structure data by Steffen (1985). 
The zirconium abundances were determined from the Zr\,{\sc i} lines at 4687.8 and 6127.5~{\AA}.
We adopted the barium hyperfine structure and isotopic composition for the 
Ba\,{\sc ii} lines at 5853.7 and 6141.7~{\AA} from
McWilliam (1998) and for the line at 6496.9~{\AA} from Mashonkina \& Gehren (2000).
Lanthanum abundances were determined from the La\,{\sc ii} lines at 6320.4 and 
6390.5~{\AA}, and cerium from the Ce\,{\sc ii} lines at 5274.2 and 6043.4~{\AA}. 
Neodymium abundances were determined using atomic parameters presented by 
Den Hartog et al.\ (2003). Due to the line crowding in the region of neodymium lines, 
only three Nd\,{\sc ii} lines were chosen: 5092.8, 5249.6 and 5319.8~{\AA}. 

   \begin{table}
\begin{center}
      \caption{Effects on derived abundances resulting from model changes 
for the star NGC\,2506\,438. The table entries show the effects on the 
logarithmic abundances relative to hydrogen, $\Delta$[El/H]. Note that the 
effects on ``relative" abundances, for example [El/Fe], are often 
considerably smaller than abundances relative to hydrogen, [El/H] } 
        \label{Sens}
      \[
         \begin{tabular}{lrrcc}
            \hline
            \noalign{\smallskip}
Species & ${ \Delta T_{\rm eff} }\atop{ \pm100 {\rm~K} }$ & 
            ${ \Delta \log g }\atop{ \pm0.3 }$ & 
            ${ \Delta v_{\rm t} }\atop{ \pm0.3 {\rm km~s}^{-1}}$ & Total\\ 
            \noalign{\smallskip}
            \hline
            \noalign{\smallskip}
C\,(C$_2$) & 0.03 & 0.03 & 0.00 & 0.04 \\
N\,(CN) & 0.03 & 0.01 & 0.02 & 0.04 \\
O\,([O\,{\sc I}]) & 0.02 & 0.03 & 0.01 & 0.04 \\
Na\,{\sc i} & 0.07 & 0.03 & 0.06 & 0.10 \\
Mg\,{\sc i} & 0.06 & 0.01 & 0.06 & 0.09 \\
Al\,{\sc i} & 0.06 & 0.01 & 0.03 & 0.07 \\
Si\,{\sc i} & 0.01 & 0.04 & 0.04 & 0.05 \\
Ca\,{\sc i} & 0.09 & 0.02 & 0.10 & 0.13 \\
Sc\,{\sc ii} & 0.01 & 0.13 & 0.08 & 0.15 \\
Ti\,{\sc i} & 0.14 & 0.01 & 0.08 & 0.16 \\
Ti\,{\sc ii} & 0.01 & 0.13 & 0.11 & 0.17 \\
V\,{\sc i} & 0.15 & 0.01 & 0.06 & 0.16 \\
Cr\,{\sc i} & 0.10 & 0.01 & 0.09 & 0.14 \\
Cr\,{\sc ii} & 0.05 & 0.12 & 0.07 & 0.15 \\
Mn\,{\sc i} & 0.08 & 0.02 & 0.10 & 0.13 \\
Fe\,{\sc i} & 0.08 & 0.00 & 0.09 & 0.12 \\
Fe\,{\sc ii} & 0.07 & 0.14 & 0.10 & 0.19 \\
Co\,{\sc i} & 0.05 & 0.02 & 0.06 & 0.08 \\
Ni\,{\sc i} & 0.03 & 0.02 & 0.08 & 0.09 \\
Cu\,{\sc i} & 0.03 & 0.02 & 0.08 & 0.09 \\
Zn\,{\sc i} & 0.02 & 0.07 & 0.08 & 0.11 \\
Y\,{\sc ii} & 0.02 & 0.10 & 0.11 & 0.15 \\
Zr\,{\sc i} & 0.04 & 0.09 & 0.04 & 0.11 \\
Ba\,{\sc ii} & 0.02 & 0.12 & 0.08 & 0.15 \\
La\,{\sc ii} & 0.08 & 0.11 & 0.02 & 0.13 \\
Ce\,{\sc ii} & 0.01 & 0.13 & 0.05 & 0.14 \\
Pr\,{\sc ii} & 0.02 & 0.13 & 0.01 & 0.13 \\
Nd\,{\sc ii} & 0.02 & 0.11 & 0.06 & 0.12 \\
Eu\,{\sc ii} & 0.00 & 0.13 & 0.02 & 0.13 \\
\\
C/N & 0.16 & 0.05 & 0.04 & 0.17 \\
$^{12}$C/$^{13}$C & 1.5 & 1.3 & 0.5 & 2 \\
                 \noalign{\smallskip}
            \hline
         \end{tabular}
      \]
  \end{center}
   \end{table}

For this cluster, along with europium, another r-process element (praseodymium) was investigated. 
The praseodymium abundances were determined from the Pr\,{\sc ii} line at 5322.8~{\AA} 
(in Fig.~6, a fit to the NGC\,2506\,438 spectrum is shown). 
Europium abundances were determined using the Eu\,{\sc ii} line at 6645.1~{\AA}. 
A hyperfine structure for the Eu\,{\sc ii} line was used for the line synthesis. 
In Fig.~7, a fit to the Eu\,{\sc ii} line in the NGC\,2506\,438 spectrum is shown. 

\subsection{Estimation of uncertainties}

The sensitivity of the abundance 
estimates to changes in the atmospheric parameters by the assumed errors 
($\pm~100$~K for $T_{\rm eff}$, $\pm 0.3$~dex for log~$g$ and 
$\pm 0.3~{\rm km~s}^{-1}$ for $v_{\rm t}$) is 
illustrated  for the star NGC\,2506\,438 (Table~2). It is seen that possible 
parameter errors do not affect the abundances seriously; the element-to-iron 
ratios, which we use in our discussion, are even less sensitive. 

The scatter of the deduced line abundances $\sigma$, presented in Table~3, 
gives an estimate of the uncertainty due to the random errors, for example, in 
continuum placement and the line parameters (the mean value of  $\sigma$ 
is 0.07). Thus the uncertainties in the derived abundances that are the 
result of random errors amount to approximately this value. 

For the giant star 456, whose spectrum has the lowest signal-to-noise ratio, we have 
tried to select lines for the analysis very carefully, so that $\sigma$ values 
are not high. However, the results should still be considered with some caution. 
The same has to be said about the RGB-tip star 459, for which the [Fe/H] value was found 
by 0.2~dex lower than for other investigated stars (Carretta et al.\ 2004). 
     
Since abundances of C, N and O are bound together by the molecular equilibrium 
in the stellar atmosphere, we have also investigated how an error in one of 
them typically affects the abundance determination of another. 
$\Delta{\rm [O/H]}=0.10$ causes 
$\Delta{\rm [C/H]}=0.05$ and $\Delta{\rm [N/H]}=-0.10$,   
$\Delta{\rm [C/H]}=0.10$ causes $\Delta{\rm [N/H]}=-0.15$ and 
$\Delta{\rm [O/H]}=0.02$, and 
$\Delta {\rm [N/H]}=0.10$ has no effect on either the carbon or the oxygen abundances.

Other sources of uncertainties were described in detail by Mikolaitis et al.\ (2010).  

\section{Results and discussion}

The abundances of different chemical elements relative to iron [El/Fe] 
and $\sigma$ (the line-to-line 
scatter) derived for up to 30 neutral and ionized 
species for the 
programme stars are listed in Table~3.
The average cluster abundances and dispersions about the mean values for NGC\,2506 are 
presented in Table~3 as well. Except carbon and nitrogen, the majority of the investigated 
chemical elements have abundance ratios close to the solar ones. 
In NGC\,2506, the mean cluster [$\alpha/{\rm Fe}] \equiv {1\over 4}
([{\rm Mg}/{\rm Fe}]+[{\rm Si}/{\rm Fe}]+[{\rm Ca}/{\rm Fe}]
+[{\rm Ti}/{\rm Fe}]) = 0.0\pm0.06$~(s.d.). 

\begin{table*}
\begin{center}
  \caption{Abundances relative to iron [El/Fe]. The quoted 
errors, $\sigma$, are the standard deviations in the mean value due to the 
line-to-line scatter within the species. The number of lines used is indicated by $n$. 
The last two columns give the mean [El/Fe] and standard deviations for the cluster stars.}
\begin{tabular}{lrrrrrrrrrrrrrrrrrrrrrrrrrr}
  \hline
\scriptsize
  & \multicolumn{3}{c}{438 (clump)} &
  & \multicolumn{3}{c}{443 (clump)} &
  & \multicolumn{3}{c}{456 (giant)} &
  & \multicolumn{3}{c}{459 (RGB-tip)}&
  & \multicolumn{2}{c}{Mean}\\
            \noalign{\smallskip}
\cline{2-4}\cline{6-8}\cline{10-12}\cline{14-16}\cline{18-19}
            \noalign{\smallskip}
Species &[El/Fe] &$\sigma$ &$n$&\ &[El/Fe] &$\sigma$ &$n$&\ &[El/Fe] &$\sigma$ &$n$&\ &[El/Fe] &$\sigma$ &$n$&\ &[El/Fe] & $\sigma$\\ 
            \noalign{\smallskip}
            \hline
            \noalign{\smallskip}

C\,(C$_2$)  	&	--0.24	&		&	1	&&	--0.27	&		&	1	&&	--0.15	&		&	1	&&	--0.10	&		&	1	&&	--0.19	&	0.08	\\
N\,(CN)     	&	0.30	&	0.09	&	12	&&	0.36	&	0.08	&	12	&&	0.30	&	0.09	&	10	&&	0.31	&	0.09	&	19	&&	0.32	&	0.03	\\
O\,([O\,{\sc I}])	&	--0.03	&		&	1	&&	0.02	&		&	1	&&	0.03	&		&	1	&&	0.15	&		&	1	&&	0.04	&	0.08	\\
Na\,{\sc i} 	&	0.01	&	0.07	&	2	&&	0.03	&	0.06	&	2	&&	--0.11	&	0.06	&	2	&&	0.14	&	0.05	&	2	&&	0.02	&	0.10	\\
Mg\,{\sc i} 	&	0.04	&	0.06	&	3	&&	0.09	&	0.06	&	3	&&	--0.01	&		&	1	&&	0.05	&	0.07	&	3	&&	0.04	&	0.04	\\
Al\,{\sc i} 	&	0.04	&	0.04	&	 2	&&	--0.01	&	0.05	&	 2	&&	--0.06	&	0.00	&	 1	&&	0.12	&	0.04	&	 2	&&	0.02	&	0.07	\\
Si\,{\sc i} 	&	--0.05	&	0.09	&	10	&&	--0.01	&	0.06	&	12	&&	--0.01	&	0.03	&	 4	&&	0.12	&	0.07	&	12	&&	0.01	&	0.07	\\
Ca\,{\sc i} 	&	--0.06	&	0.06	&	 6	&&	--0.08	&	0.06	&	 6	&&	--0.12	&	0.03	&	 3	&&	0.02	&	0.07	&	 6	&&	--0.06	&	0.06	\\
Sc\,{\sc ii}	&	--0.02	&	0.08	&	 6	&&	--0.04	&	0.08	&	 8	&&	0.10	&	0.08	&	 4	&&	0.06	&	0.09	&	 9	&&	0.03	&	0.07	\\
Ti\,{\sc i} 	&	--0.01	&	0.08	&	18	&&	--0.02	&	0.10	&	23	&&	--0.08	&	0.94	&	10	&&	0.07	&	0.09	&	25	&&	--0.01	&	0.06	\\
Ti\,{\sc ii}	&	0.00	&	0.09	&	 6	&&	--0.04	&	0.08	&	 7	&&	0.08	&	0.09	&	 4	&&	0.09	&	0.09	&	10	&&	0.03	&	0.07	\\
V\,{\sc i}  	&	--0.02	&	0.08	&	 6	&&	0.00	&	0.09	&	 7	&&	--0.11	&	0.08	&	 4	&&	0.04	&	0.09	&	 7	&&	--0.02	&	0.06	\\
Cr\,{\sc i} 	&	--0.12	&	0.09	&	17	&&	--0.11	&	0.09	&	19	&&	--0.08	&	0.09	&	 8	&&	--0.02	&	0.09	&	26	&&	--0.08	&	0.05	\\
Cr\,{\sc ii}	&	--0.03	&	0.08	&	 9	&&	0.02	&	0.06	&	 8	&&	0.13	&	0.06	&	 2	&&	--0.04	&	0.09	&	 9	&&	0.02	&	0.08	\\
Mn\,{\sc i} 	&	--0.17	&	0.08	&	 6	&&	--0.15	&	0.07	&	 4	&&	--0.17	&	0.04	&	 3	&&	0.00	&	0.09	&	 6	&&	--0.12	&	0.08	\\
Co\,{\sc i} 	&	--0.03	&	0.07	&	 5	&&	--0.03	&	0.09	&	 5	&&	0.00	&	0.02	&	 2	&&	0.09	&	0.04	&	 7	&&	0.01	&	0.06	\\
Ni\,{\sc i} 	&	--0.12	&	0.09	&	31	&&	--0.13	&	0.09	&	32	&&	--0.07	&	0.09	&	19	&&	--0.01	&	0.09	&	32	&&	--0.08	&	0.06	\\
Cu\,{\sc i} 	&	--0.13	&		&	1	&&	--0.13	&		&	1	&&	--0.14	&		&	1	&&	--0.21	&		&	1	&&	--0.15	&	0.04	\\
Zn\,{\sc i} 	&	0.01	&	0.08	&	3	&&	0.00	&	0.08	&	3	&&	0.05	&		&	1	&&	0.04	&	0.04	&	3	&&	0.02	&	0.02	\\
Y\,{\sc ii} 	&	--0.09	&	0.02	&	5	&&	0.01	&	0.07	&	5	&&	0.09	&	0.03	&	3	&&	--0.14	&	0.04	&	5	&&	--0.03	&	0.10	\\
Zr\,{\sc i} 	&	0.13	&		&	1	&&	0.13	&	0.07	&	2	&&	0.13	&		&	1	&&	--0.03	&	0.03	&	2	&&	0.09	&	0.08	\\
Ba\,{\sc ii}	&	0.08	&	0.08	&	3	&&	0.10	&	0.07	&	2	&&	0.07	&	0.05	&	3	&&	--0.11	&	0.05	&	3	&&	0.04	&	0.10	\\
La\,{\sc ii}	&	0.11	&	0.07	&	2	&&	0.13	&	0.04	&	2	&&		&		&		&&	--0.06	&	0.07	&	2	&&	0.06	&	0.10	\\
Ce\,{\sc ii}	&	0.02	&		&	1	&&	0.14	&	0.04	&	2	&&	--0.04	&		&	1	&&	0.19	&	0.04	&	2	&&	0.08	&	0.11	\\
Pr\,{\sc ii}	&	0.13	&		&	1	&&	0.19	&		&	1	&&		&		&		&&	0.14	&		&	1	&&	0.15	&	0.03	\\
Nd\,{\sc ii}	&	0.01	&	0.04	&	2	&&	0.05	&	0.04	&	2	&&	0.10	&		&	1	&&	0.19	&	0.09	&	3	&&	0.09	&	0.08	\\
Eu\,{\sc ii}	&	0.17	&	     	&	1	&&	0.24	&	      	&	1	&&	0.20	&	      	&	1	&&	0.19	&	      	&	1	&&	0.20	&	0.03	\\
\\              		      		     		   		       		      		    		      		      		  		      		      		   		      		    	
C/N             	&	1.14	&	     	&	   	&&	0.93	&	      	&	    	&&	1.39	&	      	&		&&	1.55	&	      	&	   	&&	1.25	&	0.27	\\
$^{12}$C/$^{13}$C	&	10	&	     	&	   	&&	8	&	      	&	    	&&		&	      	&	  	&&	14	&	      	&	   	&&	11	&	3	\\
             \noalign{\smallskip}   
\hline
\end{tabular}
\end{center}
\end{table*}

In the remaining of this section, we will discuss in more detail the carbon and nitrogen abundance results. 
Investigations of abundances of these chemical elements in atmospheres of clump and giant stars 
of open clusters may provide a comprehensive information on 
chemical composition changes during their evolution along the giant branch and at the helium flash.  

\input epsf
\begin{figure}
\epsfxsize=\hsize 
\epsfbox[-20 -20 620 500] {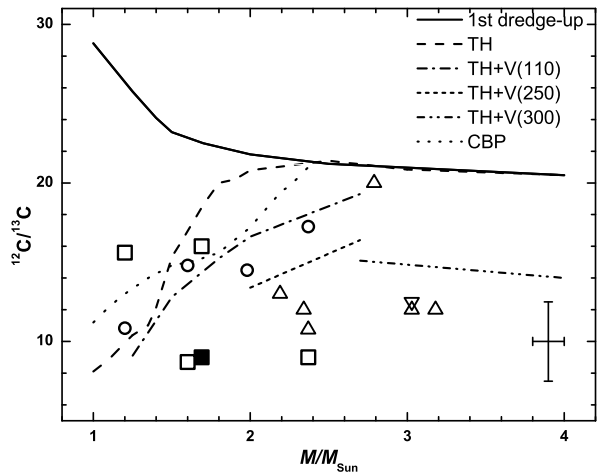} 
    \caption {The average carbon isotope ratios in clump stars of open cluster as a function of stellar turn-off mass. The models of the first dredge-up, 
thermohaline mixing (TH) and rotation-induced mixing (V) are taken from Charbonnel \& Lagarde (2010). 
The CBP model of extra-mixing is taken from Boothroyd \& Sackmann (1999). 
The result of this work is marked by the filled square; 
from Mikolaitis et al.\ (2010, 2011) and Tautvai\v{s}ien\.{e} et al.\ (2000, 2005) -- open squares; 
from Smiljanic et al.\ (2009) -- open triangles; from Luck (1994) -- reversed open triangle;
from Gilroy (1989) -- open circles. A typical error bar is indicated. 
}
    \label{Fig8}
  \end{figure}

\input epsf
\begin{figure}
\epsfxsize=\hsize 
\epsfbox[-20 -20 620 500] {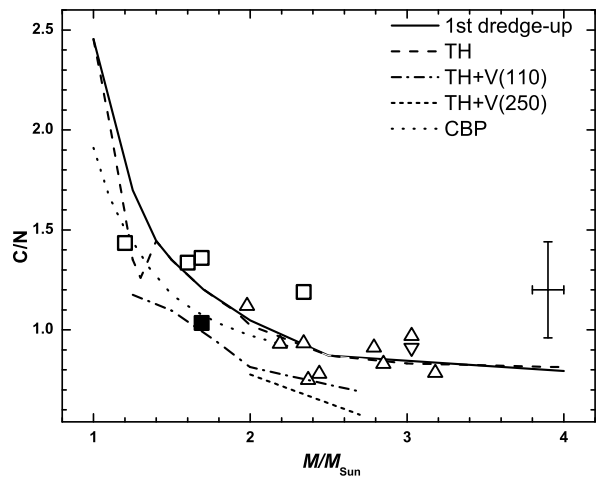} 
    \caption {The average carbon to nitrogen ratios in clump stars of open clusters as a function of stellar
              turn-off mass. The meaning of symbols are as in Fig.~8.}
    \label{Fig9}
  \end{figure}

The average value of carbon to iron ratio in NGC\,2506 is ${\rm [C/Fe]}=-0.19\pm0.08$. 
We compared the carbon abundance in NGC\,2506 with carbon abundances determined for dwarf stars 
in the Galactic disc. Shi, Zhao \& Chen (2002) performed an abundance 
analysis of carbon for a sample of 90 F type and G type main-sequence (MS) disc stars  
using C\,{\sc i} and [C\,{\sc i}] lines and found [C/Fe] to be about solar 
at the solar metallicity.   
Roughly solar carbon abundances were found by Gustafsson et al.\ (1999) 
who analysed a sample of 80 late F and early G type dwarfs using the forbidden 
[C\,{\sc i}] line. 
The ratios of [C/Fe] in our stars lie about 0.2~dex below 
the values obtained for dwarf stars of the Galactic disc.      

The mean nitrogen to iron abundance ratio 
in NGC\,2506 is ${\rm [N/Fe]}=0.32\pm0.03$. 
This shows that nitrogen is overabundant in these evolved stars of NGC\,2506 by 0.3~dex,
since [N/Fe] values in the Galactic MS stars are about solar at the solar
metallicity (c.f. Shi, Zhao \& Chen 2002).

$^{12}{\rm C}/^{13}{\rm C}$ and C/N ratios may provide an important information on mixing processes in stars. 
The mean $^{12}{\rm C}/^{13}{\rm C}$ and C/N ratios in the investigated NGC\,2506 stars are equal to $11\pm 3$ and 
$1.25\pm 0.27$, respectively. 
In Mikolaitis et al.\ (2011) we have compiled recent data on $^{12}{\rm C}/^{13}{\rm C}$ and C/N ratios 
in clump stars of open clusters. The clump stars have accumulated all chemical composition changes that have 
happened during their evolution along the giant branch, so they are very useful indicators of abundance alterations. 

In Fig.~8 and 9, we compare the mean $^{12}{\rm C}/^{13}{\rm C}$ and C/N ratios 
of cluster clump stars as 
a function of turn-off mass with the theoretical models of the $1^{st}$ dredge-up, thermohaline mixing (TH), 
TH together with rotation-induced mixing for stars at the zero age main sequence having 
rotational velocities of 110, 250 and 300\,km\,s$^{-1}$ computed by Charbonnel \& Lagarde (2010), 
Cool Bottom Processing (CBP) model by Boothroyd \& Sackman (1999)  and previous investigations 
of $^{12}{\rm C}/^{13}{\rm C}$ and C/N ratios by Mikolaitis et al.\ (2010, 2011), Tautvai\v{s}ien\.{e} et 
al.\ (2000, 2005), Smiljanic et al.\ (2009), Luck (1994) and Gilroy (1989). The turn-off mass of stars in 
NGC\,2506 was taken from C04. The typical error bars are indicated as well (c.f. Charbonnel \& Lagarde 2010).

The $^{12}{\rm C}/^{13}{\rm C}$ values we derive for the stars in the open cluster NGC\,2506 confirm 
the observational evidence that theoretical models for stars with larger turn-off masses should consider 
the larger extra-mixing, probably dominated by the former rotation on the main sequence (Mikolaitis et al.\ 2011; 
Charbonnel \& Lagarde 2010). 
 
In NGC\,2506, the mean C/N ratios in the clump stars are lowered slightly more 
($1.04\pm 0.11$) than in the first ascent giants ($1.47\pm 0.08$). This reminds us of an unanswered question about 
the He-flash influence to mixing processes of CN-processed material in giants. In the clump stars the 
$^{12}{\rm C}/^{13}{\rm C}$ ratios are also lowered to the smaller values (8 and 10) than in the RGB-tip star 459, 
which is equal to 14. The differences of $^{12}{\rm C}/^{13}{\rm C}$ and C/N ratios in the clump stars and giants might 
be caused by the He-flash.  

The He-flash influence to the extra-mixing of CN-cycled material to stellar surfaces still has to be 
investigated both theoretically and observationally. The theoretical 
calculations indicate that the nature of nucleosynthesis and mixing depend 
upon the degree of degeneracy in the He-core and, hence, intensity of the 
explosion: intermediate flashes produce most efficient mixing (Despain 1982; Deupree 1982; Deupree \& Wallace 1987). 
Attempts to model this violent event of stellar evolution are continuing (e.g. Schlattl et al.\ 2001; 
Cassisi et al.\ 2003; Dearborn, Lattanzio \& Eggleton 2006; Moc\'{a}k et al.\ 2010 and references therein).

Precise observations of RGB-tip stars 
and clump stars in clusters are most useful in order to uncover effects of the 
He-core flash. In the previously investigated open clusters M\,67 and NGC\,7789 we 
also found some differences in the mean $^{12}{\rm C}/^{13}{\rm C}$ and $^{12}{\rm C}/^{14}{\rm N}$ 
ratios when comparing giants and clump stars (Tautvai\v{s}ien\.{e} et al. 
2000, 2005).  
In M\,67, with a mass of turn-off stars of about 1.2~$M_{\odot}$, the mean 
$^{12}{\rm C}/^{13}{\rm C}$ in giants is lowered to the value 
of $24 \pm 4$, and the $^{12}{\rm C}/^{14}{\rm N}$ ratio to the value of $1.7 \pm 0.2$. 
In the clump stars observed the mean are lower: $^{12}{\rm C}/^{13}{\rm C}=16\pm 4$ 
and $^{12}{\rm C}/^{14}{\rm N}=1.4\pm 0.2$. In NGC\,7789 we also investigated the 
first-ascent giants located above the red giant bump and more evolved clump stars.
The mass of turn-off stars is of about 1.6~$M_{\odot}$. The mean 
$^{12}{\rm C}/^{14}{\rm N}$ ratio is $1.9\pm0.5$ in the giants, and $1.3\pm 0.2$ in the clump stars;    
however, the $^{12}{\rm C}/^{13}{\rm C}$ ratios are very similar for 
all the stars investigated, $9\pm 1$. A RGB-tip star and two clump stars were 
investigated in the cluster NGC\,3532 by Smiljanic et al.\ (2009). The mean
$^{12}{\rm C}/^{13}{\rm C}$ ratio in the clump stars is $11\pm1$, while in the RGB-tip 
star is larger and equal to 20. However, the C/N ratios are about the same. 
Three clump stars and one RGB-tip star were investigated in the open cluster IC\, 4561 
by Mikolaitis et al. (2011).  Due to the different [Fe/H], which is by 0.4~dex lower than of other cluster 
stars, the RGB-tip star is not trustful enough for the investigations of possible tinny effects of He-core flash. 

$^{12}{\rm C}/^{13}{\rm C}$ ratios little depend on stellar atmospheric parameters 
and are sensitive indicators of mixing processes, their analysis has to be continued. 
We plan to address this topic in our forthcoming papers.

\section*{Acknowledgments}

This research has made use of SIMBAD, VALD and NASA ADS databases.
Bertrand Plez (University of Montpellier II) and Guillermo Gonzalez  
(Washington State University) were particularly generous in providing us with 
atomic data for CN and C$_2$ molecules, respectively.
\v{S}.\,M. and G.\,T. were supported by the Ministry of Education and Science of Lithuania 
via LitGrid programme and by the European Commission via FP7 Baltic Grid II project. 



\begin{thebibliography}{}
 \bibitem [\protect\citeauthoryear{Alonso et al.}{2001}]{Alonso2001} Alonso A., Arribas S., Mart{\'{\i}}nez-Roger C., 2001, A\&A, 376, 1039 
 \bibitem [\protect\citeauthoryear{Boothroyd et al.}{1999}]{Boothroyd1999} Boothroyd A. I., Sackmann I. J., 1999, ApJ, 510, 232
 \bibitem [\protect\citeauthoryear{Bragaglia \& Tosi}{2006}]{Bragaglia2006} Bragaglia A., Tosi M. 2006, AJ, 131, 1544
 \bibitem [\protect\citeauthoryear{Carraro \& Chiosi}{1994}]{Carraro1994} Carraro G., Chiosi C., 1994, A\&A, 288, 751
 \bibitem [\protect\citeauthoryear{Carretta, Bragaglia \& Gratton}{2007}]{Carretta2007} Carretta  E., Bragaglia A., Gratton R. 2007, A\&A, 473, 129
 \bibitem [\protect\citeauthoryear{Carretta et al.}{2004}]{Carretta2004} Carretta  E., Bragaglia A., Gratton R., Tosi M., 2004, A\&A, 422, 951 (C04)
 \bibitem [\protect\citeauthoryear{Cassisi et al.}{2003}]{Cassis2003} Cassisi S., Schlattl H., Salaris M., Weiss A., 2003, ApJ, 582, L43 
 \bibitem [\protect\citeauthoryear{Charbonnel \& Lagarde}{2010}]{Charbonnel2010} Charbonnel C. \& Lagarde N., 2010, A\&A, 522, 10
 \bibitem [\protect\citeauthoryear{Chen, Chen \& Shu}{2004}]{Chen2004} Chen W.P., Chen C.W., Shu C.G., 2004, AJ, 128, 2306 
 \bibitem [\protect\citeauthoryear{Chiu \& van Altena}{1981}]{Chiu1981} Chiu L.-T. G., van Altena W. F., 1981, ApJ, 243, 827
 \bibitem [\protect\citeauthoryear{Dearborn, Lattanzio \& Eggleton}{2006}]{Dearborn2006} Dearborn D. S. P., Lattanzio J. C., Eggleton P. P. 2006, ApJ, 639, 405
 \bibitem [\protect\citeauthoryear{Den Hartog et al.}{2003}]{DenHartog2003} Den Hartog E. A., Lawler J. E., Sneden C., Cowan J. J., 2003, ApJS, 148, 543
 \bibitem [\protect\citeauthoryear{Despain}{1982}]{Despain1982} Despain K. H. 1982, ApJ, 253, 811
 \bibitem [\protect\citeauthoryear{Deupree}{1986}]{Deupree1986} Deupree R.G. 1986, ApJ, 303, 649
 \bibitem [\protect\citeauthoryear{Deupree \& Wallace}{1987}]{Deupree1987} Deupree R.G., Wallace R. K. 1987, ApJ, 317, 724
 \bibitem [\protect\citeauthoryear{Dias et al.}{2002}]{Dias2002} Dias W. S., Alessi B. S., Moitinho A., L\'{e}pine J. R. D., 2002, A\&A, 389, 871
 \bibitem [\protect\citeauthoryear{Geisler}{1984}]{Geisler1984} Geisler D., 1984, ApJ, 287, L85 
 \bibitem [\protect\citeauthoryear{Geisler, Claria \& Minniti}{1992}]{Geisler1992} Geisler D., Claria J. J., Minniti D., 1992, AJ, 104, 1892
 \bibitem [\protect\citeauthoryear{Gilroy et al.}{1989}]{Gilroy1989} Gilroy K. K., 1989, ApJ, 347, 835 
 \bibitem [\protect\citeauthoryear{Gratton et al.}{1999}]{Gratton1999} Gratton R.G., Carretta, E., Eriksson, K., Gustafsson, B., 1999, A\&A, 350, 955
 \bibitem [\protect\citeauthoryear{Grevesse et al.}{2000}]{Grevesse2000} Grevesse N., Sauval A.J., 2000, ``Origin of Elements in the Solar System, Implications of Post-1957 Observations, O. Manuel (ed.), Kluwer, 261
 \bibitem [\protect\citeauthoryear{Gustafsson et al.}{1999}]{Gustafsson1999} Gustafsson B., Karlsson T., Olsson E., Edvardsson B., Ryde N., 1999, A\&A, 342, 426
 \bibitem [\protect\citeauthoryear{Johansson et al.}{2003}]{Johansson2003} Johansson S., Litz\'{e}n U., Lundberg H., Zhang Z., 2003, ApJ, 584, 107
 \bibitem [\protect\citeauthoryear{Kim et al.}{2001}]{Kim2001} Kim S. L., Chun M. Y., Park B.-G., et al., 2001, AcA, 51, 49
 \bibitem [\protect\citeauthoryear{Kurucz et al.}{1984}]{Kurucz1984} Kurucz R. L., Furenlid I., Brault J., Testerman L., 1984, "Solar Flux Atlas from 296 to 1300 nm.", National Solar Observatory, Sunspot, New Mexico 
 \bibitem [\protect\citeauthoryear{Luck et al.}{1994}]{Luck1994} Luck R. E., 1994, ApJS, 91, 309
 \bibitem [\protect\citeauthoryear{Marconi et al.}{1997}]{Marconi1997} Marconi G., Hamilton D., Tosi M., Bragaglia A., 1997, MNRAS, 291, 763
 \bibitem [\protect\citeauthoryear{Mashonkina \& Gehren}{2000}]{Mashonkina2000} Mashonkina L. \& Gehren T, 2000, A\&A, 364, 249
 \bibitem [\protect\citeauthoryear{McClure, Twarog \& Forrester}{1981}]{McClure1981} McClure R. D., Twarog B. A., Forrester W. T., 1981, ApJ, 243, 841
 \bibitem [\protect\citeauthoryear{McWilliam}{1998}]{McWilliam1998} McWilliam A. 1998, AJ, 115, 1640
 \bibitem [\protect\citeauthoryear{Mikolaitis et al.}{2010}]{Mikolaitis2010} Mikolaitis \v{S}., Tautvai\v{s}ien\.{e} G., Gratton R., Bragaglia A., Carretta E., 2010, MNRAS, 407, 1866
 \bibitem [\protect\citeauthoryear{Mikolaitis et al.}{2011}]{Mikolaitis2011} Mikolaitis \v{S}., Tautvai\v{s}ien\.{e} G., Gratton R., Bragaglia A., Carretta E., 2011, MNRAS, 413, 2199 
 \bibitem [\protect\citeauthoryear{Moc\'{a}k et al.}{2010}]{Mocak2010} Moc\'{a}k M., Campbell S. W., M\"{u}ller E., Kifonidis K. 2010, A\&A, 520, 114
 \bibitem [\protect\citeauthoryear{Piatti, Claria \& Abadi}{1995}]{Piatti1995} Piatti A. E., Claria J. J., Abadi M. G., 1995, AJ, 110, 2813 
 \bibitem [\protect\citeauthoryear{Purgathofer}{1964}]{Purgathofer1964} Purgathofer A., 1964, Zeitschrift für Astrophysik, 59, 79
 \bibitem [\protect\citeauthoryear{Schlattl et al.}{2001}]{Schlattl2001} Schlattl H., Cassisi S., Salaris M., Weiss A. 2001, ApJ, 559, 1082
 \bibitem [\protect\citeauthoryear{Schlegel, Finkbeiner, \& Davis}{1998}]{Schlegel1998} Schlegel D. J., Finkbeiner D. P., Davis M., 1998, ApJ, 500, 525 
 \bibitem [\protect\citeauthoryear{Shi, Zhao \& Chen}{2002}]{Shi2002} Shi J. R., Zhao G., Chen Y. Q., 2002, A\&A, 381, 982
 \bibitem [\protect\citeauthoryear{Smiljanic et al.}{2009}]{Smiljanic2009} Smiljanic R., Gauderon R., North P., Barbuy, B., Charbonnel C., Mowlavi N., 2009, A\&A, 502, 267
 \bibitem [\protect\citeauthoryear{Steffen}{1985}]{Steffen1985} Steffen M., 1985, A\&AS, 59, 403  
 \bibitem [\protect\citeauthoryear{Tautvai\v{s}ien\.{e} et al.}{2000}]{Tautvaisiene2000} Tautvai\v{s}ien\.{e} G. Edvardsson B., Tuominen I., Ilyin I., 2000, A\&A, 360, 499
 \bibitem [\protect\citeauthoryear{Tautvai\v{s}ien\.{e} et al.}{2005}]{Tautvaisiene2005} Tautvai\v{s}ien\.{e} G. Edvardsson B., Puzeras E., Ilyin I., 2005, A\&A, 431, 933
 \bibitem [\protect\citeauthoryear{van der Bergh \& Sher}{1960}]{vandenBergh1960} van den Bergh S., Sher D., 1960, Pub. David Dunlap Obs., 2, 203 
\end{thebibliography}
\end{document}